\documentclass[aps,prl,twocolumn,letterpaper,showpacs,preprintnumbers,amsmath,amssymb]{revtex4}

\usepackage{epsfig,amssymb}
\usepackage{graphicx}
\usepackage{dcolumn}
\usepackage{bm}
\usepackage{dcolumn}
\usepackage{multirow}

\begin{document}

\title{Material limitations of carbon-nanotube inertial balances:
on the possibility of intrinsic yoctogram mass resolution at room temperature}

\author{Johannes~Lischner}
\author{T.A.~Arias}
\affiliation{Laboratory of Atomic and Solid State Physics, Cornell University, Ithaca, New
York 14853}

\begin{abstract}
We present a theoretical study of the intrinsic quality factor of the
fundamental flexural vibration in a carbon nanotube and its dependence
on temperature, radius, length and tension. In particular, we examine
three- and four-phonon decays of the fundamental flexural mode within
quantized elasticity theory. This analysis reveals design principles
for the construction of ultrasensitive nanotube mass sensors: under
tensions close to the elastic limit, intrinsic losses allow for
\emph{single yoctogram} mass resolution at room temperature, while
cooling opens the possibility of \emph{sub-yoctogram} mass resolution.

\end{abstract}
\maketitle

Recently, much experimental effort has gone into fabricating
nano-electromechanical systems (NEMS) which employ carbon nanotubes as
mechanical resonators \cite{Bachtold2,Steele,vanderZant,Zettl}. The
combination of small mass density and large mechanical stiffness makes
carbon nanotube NEMS ideal candidates for ultrasensitive mass
detectors which do not require disruptive ionization of the analyzed
molecule: recent experiments achieved a mass resolution of $\sim
100$~yg ($1$~yg=$10^{-24}$~g)\cite{Bachtold,Bockrath}.  However, for
isotopic or chemical identification of molecules adsorbed on the
nanotube or real-time monitoring of chemical reactions
\emph{yoctogram} mass resolution is necessary.

The mass resolution of nanotube NEMS depends strongly on the quality
factor of the lowest flexural mode, which imposes a lower bound on the
frequency differences which can be resolved. The quality factors found
in most experiments have been surprisingly low, not exceeding $Q
\approx 2000$ \cite{Bachtold,Zettl,vanderZantBending}. Only very
recently, H\"uttel et al. \cite{vanderZant} measured $Q \gtrsim10^5$,
but at {\em milikelvin} temperatures.

The \emph{intrinsic} quality factor of a nanotube sets aside all
extrinsic losses, such as defect or clamping losses, and sets an upper
limit to the quality factor achievable in experiment. Both molecular
dynamics simulations \cite{Jiang} and analytical approaches
\cite{Srivastava,DeMartino} have been used to study the intrinsic
quality factor theoretically, finding relatively low values similar to
those in the experiments. However, simulations are limited to
extremely short tubes, and the analytical approaches to date have
either employed simplified phonon spectra\cite{Srivastava} or focused
solely on low-temperature four-phonon decays of the flexural
mode\cite{DeMartino} using an analysis appropriate to extremely long
tubes.

In this Letter, we present a theoretical analysis appropriate to the
length of tubes and operating temperatures commonly found in
experiments.  We discover that application of tension drastically
increases the intrinsic quality factor of nanotube oscillators and
opens for the first time the theoretical possibility of \emph{single
  yoctogram} mass resolution in such oscillators at room temperature,
with further improvements possible with cooling.  Indeed, the recent
experiments of Wei et al. demonstrate a promising technique for
controlling the tension in nanotubes\cite{Wei}.

For our analysis, we employ continuum elastic theory, which reliably
describes long wavelength phonons in nanotubes, to study the decay of
the lowest flexural mode due to phonon-phonon interactions, the most
important source of intrinsic losses in semiconducting, and possibly
also metallic, nanotubes \cite{DeMartino}. Following the work of De
Martino et al. \cite{DeMartino} and Suzuura et al. \cite{Ando}, we
describe a nanotube as a rolled-up two dimensional elastic sheet and
expand the free energy in powers of the strain tensor $u_{ij}$ and the
curvature tensor $S_{ij}$. Curvature contributions to the free energy
are generally small, but need to be included in the quadratic part to
describe optical phonons properly. Using the isotropy of the elastic
sheet leads to
\begin{eqnarray}
{\cal U}_2 &=& \frac{\lambda}{2} ({\rm Tr} u)^2+\mu {\rm Tr} u^2
+\frac{k_c}{2} \left( {\rm Tr}(S-S_0) \right)^2, \label{U2}\\
{\cal U}_3 &=& \zeta_1 ({\rm Tr}u)^3 +
\zeta_2 {\rm Tr}u^2 {\rm Tr}u, \label{U3}
\end{eqnarray}
where ${\cal U}_2$ and ${\cal U}_3$ denote contributions to the energy
density including quadratic and cubic powers of $u_{ij}$ and
$S_{ij}$. Also, $\lambda$ and $\mu$ denote the Lam\'e moduli, $k_c$
the bending stiffness and $S_0$ the curvature tensor of the undeformed
tube. $\zeta_1$ and $\zeta_2$ are anharmonic elastic constants.

Next, using the standard expressions for $u_{ij}$ and $S_{ij}$ in
cylindrical coordinates \cite{DeMartino,Ando}, we express the
potential energy in terms of displacements $u_n$ with $n=x,y,z$ ($x$
denotes the tangential direction, $y$ is along the tube axis and $z$
is the radial direction). $u_{ij}$ contains both linear and nonlinear
contributions in $u_n$, which gives rise to two types of
nonlinearities in the resulting Hamiltonian: material nonlinearities,
described by the explicitly higher-order terms in Eq.~(\ref{U3}), and
geometric nonlinearities, where the lower order terms, such as terms
in Eq.~(\ref{U2}), contribute at higher orders due to higher order
terms in the expansion of $u_{ij}$ in terms of the degrees of freedom
$u_n$.  Material nonlinearities are often suppressed by a factor of
$kR$ ($k$ being the phonon wave vector and $R$ the tube radius)
compared to geometric nonlinearities \cite{DeMartino} and can
therefore be neglected for long wavelength phonons. However, when
evaluating the three-phonon contribution to the quality factor, high
energy phonons, whose wavelength is no longer comparable to the length
of the tube, play a crucial role. We, therefore, also consider in this
work the contribution of material nonlinearities in the three-phonon
decay channel.

To determine accurate numerical values of the elastic constants in
Eqs.~(\ref{U2}) and (\ref{U3}), we evaluate total energies of strained
graphene sheets and various semiconducting nanotubes within
density-functional theory in the local density approximation
\cite{KohnSham} and fitted the results to the continuum theory
described above.  Table~\ref{CompareElastic} summarizes our results,
which for the elastic constants of ${\cal U}_2$ are in good agreement
with previous calculations \cite{Katsnelson,Rubio}.

\begin{table}
  \setlength{\doublerulesep}{1\doublerulesep}
  \setlength{\tabcolsep}{3\tabcolsep}
  \begin{ruledtabular}
    \begin{tabular}{c c c c c}
      $\lambda$ [J$/$m$^2$] & $\mu$ [J$/$m$^2$] & $\zeta_1$ [J$/$m$^2$] & $\zeta_2$ [J$/$m$^2$] & $k_c$ [eV]\\
      \hline
      & & & & \\
    59.57 & 147.94 & -145.17 & -387.93 & 1.46\\
    \end{tabular}
  \end{ruledtabular}
  \caption{Lam\'e moduli, anharmonic elastic constants of ${\cal U}_3$
    and bending stiffness for semiconducting carbon nanotubes from
    \emph{ab initio} density-functional calculations.}
  \label{CompareElastic}
\end{table}

We next impose canonical commutation relations to quantize the
resulting elastic theory and diagonalize the quadratic part of the
Hamiltonian by expressing the displacement operator in terms of normal
modes.

\begin{figure}
  \includegraphics[width=8.cm]{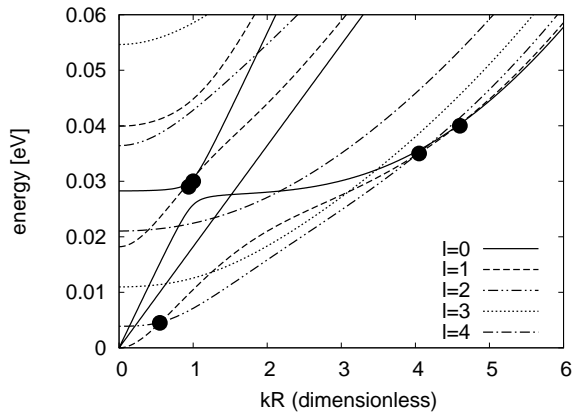}
  \caption{Energy of phonons versus $kR$ for a $R=0.5$~nm tube. The
    black dots denote phonon crossings, where three-phonon decays of
    the fundamental flexural mode are allowed.}
\label{Dispersion}
\end{figure}

Figure~\ref{Dispersion} shows the resulting phonon dispersion curves
for a tube with radius $R=0.5$~nm, which are in good agreement with
force constant models \cite{Galli} and \emph{ab initio} calculations
\cite{Rubio}.  In particular, we find as the lowest-frequency long
wavelength modes two degenerate flexural modes with angular momentum
component $\ell=\pm 1$ along the tube axis and a quadratic dispersion
relation
\begin{equation}
\omega_F(k)=\frac{\hbar k^2}{2m}+{\cal O}(k^4),  \;\;\;\;\;
m=\frac{\hbar}{R}\sqrt{\frac{\rho_M(\lambda+2\mu)}{8\mu(\lambda+\mu)}},
\label{wF}
\end{equation}
where $\omega_F$ denotes the angular frequency of the flexural phonon
and $\rho_M=7.68\times 10^{-7}$~kg/m$^2$ is the areal mass density of
graphene. We also find a variety of low-lying optical phonons. Note
that inclusion of curvature terms into Eq.~(\ref{U2}) is crucial to
obtain the correct optical phonon spectrum. In particular, the gap of
the lowest optical branch vanishes if curvature terms are neglected.

Next, we employ conservation laws to analyze the possible decay
mechanisms of the fundamental flexural mode. In addition to energy
conservation, translational invariance along the tube axis imposes
conservation of $k$. Rotational invariance around the tube axis
imposes conservation of $\ell$. These conservation laws forbid the
decay of the fundamental flexural mode in a three-phonon process which
involves two other low-frequency modes \cite{DeMartino}. If, however,
the fundamental flexural mode, which has a wavelength comparable to
the length of the tube, interacts with a high-energy phonon of
momentum $k_{\mu}$, energy and momentum conservation enforce that the
third phonon also has a high energy and momentum $k_{\nu} \approx
k_{\mu}$. Because the flexural mode carries $\ell=\pm 1$, the angular
momenta of the two high-energy phonons must differ by one. In sum,
three-phonon decays of the fundamental flexural mode are only possible
at crossings of two phonon bands whose angular momentum quantum
numbers differ by one. Inspection of Fig.~\ref{Dispersion} reveals
that very few such crossings for energies comparable and smaller than
$k_BT\approx 24$~meV (at room temperature) exist.

To compute the three-phonon contribution to the quality factor, we
evaluate the imaginary part of the Matsubara Green function obtained
from the lowest order bubble diagram, which contains a sum over
intermediate high-energy phonon momenta. Because the energy
uncertainty resulting from the short lifetimes \cite{Heinz} of these
modes is larger than the energy difference of phonons at neighboring
allowed wave vectors of the finite-length tube, the sum can be
converted into an integral. Then we use the energy-conserving
$\delta$-function to reduce the integral into a sum over allowed
crossings.  Our final expression for the inverse quality factor
resulting from the decay of a long wavelength flexural mode of wave
vector $k$ in a three-phonon process involving two high-energy modes
(labeled $\mu$ and $\nu$) at a phonon crossing at wave vector
$k_{\times}$ and angular frequency $\omega_{\times}$ is

\begin{equation}
Q^{-1}_3=
\sum_{\times} g_{F\mu\nu}(k,k_{\times},\ell_{\mu})
\frac{\beta\hbar^2 n(\omega_{\times})[n(\omega_{\times})+1]}
{R \omega_F(k) \omega^2_{\times} |v^{\times}_{\mu}-v^{\times}_{\nu}|},
\label{Qinvfinal}
\end{equation}
where $g_{F\mu\nu}$ is a complicated coupling function that depends on
the polarization vectors of all three phonons and gives a complete
description of both material and geometric nonlinearities. Also,
$\sum_{\times}$ denotes a sum over allowed crossings;
$v^{\times}_{\mu/\nu}$ are the phonon group velocities and
$n(\omega)=1/(\exp(\beta\hbar\omega)-1)$ is the Bose-Einstein factor
with $\beta=1/(k_BT)$ being the inverse thermal energy.

Experiments with doubly clamped tubes inevitably involve some amount
of strain.  Here we consider the case of positive strain,
corresponding to some amount of tension in the tube.  To study the
quality factor of such a strained tube, we expand the displacement
around the new equilibrium value, taking into account the relaxation
in the equilibrium radius. Evaluation of Eq.~(\ref{U3}) at the
strained equilibrium configuration leads to two categories of
additional contributions to the \emph{quadratic} hamiltonian: one set
of terms is already present in ${\cal U}_2$ and can be absorbed into a
redefinition of the linear elastic constants; the other set of terms
gives the expected Hamiltonian for a string under tension,
proportional to $(\partial_yu_x)^2+(\partial_yu_z)^2$ with a prefactor
proportional to the tension.

Computationally, having obtained the phonon dispersions for a given
radius, we numerically determine $k_{\times}$, $\omega_{\times}$ and
$v^{\times}_{\mu/\nu}$, which are needed to evaluate $g_{F\mu\nu}$ and
ultimately the losses, for all relevant crossings. Our results
indicate that geometric nonlinearities give the largest contribution
to three-phonon losses, with material nonlinearities contributing only
about one percent.  Also, we find that the primary effect from the
application of tension is to change the frequencies of the acoustic
phonons, while the polarization vectors and the optical phonon
frequencies change very little.

\begin{figure}
  \includegraphics[width=8.cm]{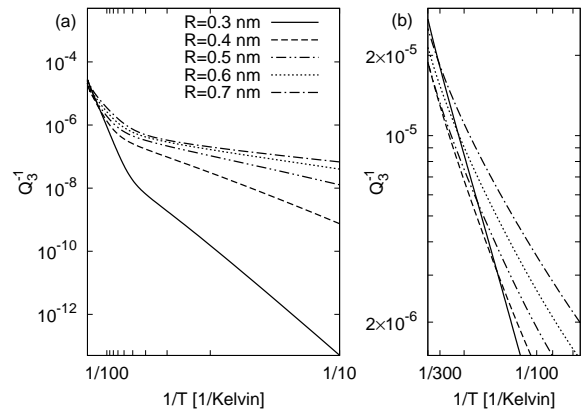}
  \caption{Log-log plots of the temperature dependence of $Q_3$ for tubes of
    different radii and length $L=500$~nm. }
\label{Tdep}
\end{figure}

Figure~\ref{Tdep} shows our results for the three-phonon contribution
to the inverse quality factors for tubes of typical experimental radii
as a function of inverse temperature. At low temperatures [right side
  of Fig.~\ref{Tdep}(a)], $Q_3^{-1}$ approaches zero exponentially,
because the first Bose-Einstein factor in Eq.~(\ref{Qinvfinal})
rapidly diminishes the occupation of the high-energy modes at the
crossings, which are the modes responsible for the scattering. At higher
temperatures, $T > 100$~K, the modes associated with the relevant
crossings are classically occupied and we find that $Q_3^{-1}$ is
proportional to temperature.

Comparing the inverse quality factors of tubes of different radii in
Fig.~\ref{Tdep}(a), we find that at low temperatures the tube with the
smallest radius, $R=0.3$~nm, exhibits the lowest dissipation, while at
high temperatures its losses are largest [see Fig.~\ref{Tdep}(b)].

To understand this nontrivial radius dependence of $Q_3$, we note that
for $T<40$~K the largest contribution to the losses comes from the
crossing which is lowest in energy. The energy of this crossing of the
flexural mode and the lowest optical mode is approximately equal to
the gap of the optical mode given by $E_{gap}=\sqrt{k_c/(5\rho_M)}
6\hbar/R^2$, which depends sensitively on the radius of the
tube. Therefore, at low temperatures, the occupation of the
high-energy modes at this crossing is much more strongly suppressed
for tubes with smaller radii (and therefore higher optical frequency),
which leads to smaller losses according to Eq.~(\ref{Qinvfinal}). At
higher temperatures, the contributions from other crossings become
important (see Fig.~\ref{Dispersion}).  We find that those
contributions depend quite sensitively on $R$, which leads to the
observed crossover behavior of the $R=0.3$~nm tube.

\begin{figure}
  \includegraphics[width=8.cm]{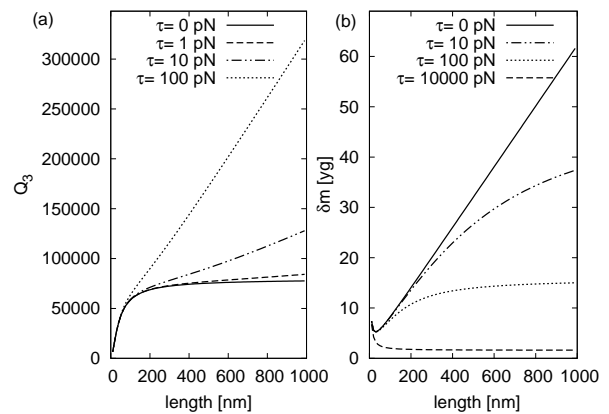}
  \caption{(a) Length dependence of $Q_3$ for tubes with $R=0.5$~nm at
    $T=300$~K. (b) Length dependence of mass resolution $\delta m$ for
    tubes with $R=0.5$~nm at $T=300$~K. }
\label{Ldep}
\end{figure}

Figure~\ref{Ldep}(a) shows the dependence of $Q_3$ on the length $L$
of the nanotube. In a tensionless tube (solid line), we find a
remarkable cancellation between the length dependence of the coupling
function, $g_{F\mu\nu} \propto k^2$ for small $k$, and the length
dependence of $\omega_F \propto k^2$ in Eq.~(\ref{Qinvfinal}),
resulting in a quality factor which is insensitive to tube length
beyond $\sim 300$~nm. Strained tubes do not exhibit this cancellation
because $\omega_F(k)$ is shifted by a constant proportional to the
tension $\tau$ if the tension is small (dashed line)\cite{Sapmaz}. If
$\tau$ is large, $\omega_F(k) \propto \sqrt{\tau} k$ and $Q_3$ becomes
linear in $L$ for long tubes (dotted line)\cite{Sapmaz}.

For tubes shorter than $300$~nm, Fig.~\ref{Ldep}(a) shows a
significant length dependence of $Q_3$ even for tensionless tubes.  In
particular, in tubes with lengths of only a few nanometers $Q_3$ is
reduced by more than an order of magnitude. This may be related to the
small quality factors, $Q\approx 1500$, found by Jiang et
al. \cite{Jiang}, who model the decay of the fundamental flexural mode
in a $3$~nm long singly clamped tube via molecular dynamics
simulations.

Next, we compare the magnitude of the computed intrinsic quality
factor to experimental findings. At $T = 300$~K, we find $Q_3\gtrsim
5\times 10^4$, which is at least one order of magnitude larger than
experimental results, $Q_{exp} \lesssim 2000$
\cite{vanderZantBending,Zettl}, suggesting that it is worthwhile to
continue improving the control of losses in experiments.  We find that
the resulting (intrinsic) mass resolution, $\delta m=2M/Q$, of a {\em
  tensionless} nanotube mass sensor depends sensitively on the tube
length $L$ with a minimum of $\sim 5$~yg for very short tubes
[Fig.~\ref{Ldep}(b)].  On the other hand, we find (i) that application
of tension can reduce the mass resolution $\delta m$ to a \emph{single
  yoctogram} if a tension close to the elastic limit, $\tau_c\approx
100$~nN \cite{Lammert}, is applied and (ii) that $\delta m$ becomes
independent of $L$ for long tubes, simplifying the design and
fabrication of actual devices.

At lower temperatures, as discussed above, three-phonon processes are
exponentially suppressed due to the energy gap of the optical modes,
and the resulting dissipation becomes much smaller than the
experimental findings in the milikelvin range \cite{vanderZant}. We,
therefore, now move on to consider losses from four-phonon decays.

To estimate the role of four-phonon processes, we compute the leading
order contribution, the fishbone diagram, due to a quartic coupling
between four low-energy flexural modes. Following De Martino et
al. \cite{DeMartino} and our findings for the three-phonon case, we
only take into account quartic geometric nonlinearities resulting from
replacing $u_{ij}$ in Eq.~(\ref{U2}) by its nonlinear part. The
resulting expression for the four-phonon contribution to the quality
factor contains a triple sum over intermediate phonon momenta and is
given by \cite{Perrin} 
\begin{eqnarray}
Q^{-1}_4&=&\sum_{q_1q_2q_3}\sum_{\xi_1,\xi_2,\xi_3=\pm}
\frac{\xi_1\xi_2\xi_3 {\cal D}(q_1,q_2,q_3,k) \eta/\pi}
{(\omega_F(k)+\xi_1\omega_1+\xi_2\omega_2+\xi_2\omega_3)^2+\eta^2} \nonumber\\
&&\times\frac{n(\xi_1\omega_1)n(\xi_2\omega_2)n(\xi_3\omega_3)}
{n(\xi_1\omega_1+\xi_2\omega_2+\xi_3\omega_3)},
\label{Qinv4}
\end{eqnarray}
where $q_i$ denote intermediate momenta, $\omega_i\equiv
\omega_F(q_i)$ and ${\cal D}$ denotes the coupling function. Here,
$\eta$ is the inverse lifetime associated with the dressed phonon
lines representing the actual phonons in the system with loss.

In evaluating Eq.~(\ref{Qinv4}), proper account of the finite length
of the tube, which leads to a finite spacing of the allowed wave
vectors, is of crucial importance. For long wavelength flexural
phonons in tubes of experimental lengths, the energy uncertainty
$\Delta E_{\eta}$ associated with the observed lifetimes is actually
much smaller than the energy difference $\Delta E_{\Delta k}$ of
phonons at neighboring wave vectors: Expressing $\eta$ in terms of the
quality factor, $\eta=\omega_F/(2\pi Q)$, we find $\Delta
E_{\eta}/\Delta E_{\Delta k}=1/(4\pi Q)$, which is much smaller than
unity at low temperatures \cite{vanderZant}. Thus, the sums over
intermediate momenta \emph{cannot} be converted into integrals for the
tubes in the experiments.  Converting the sums into integrals, as De
Martino et al.\cite{DeMartino} do, is appropriate for much longer
tubes, but leads to an underestimate of $Q_4$ due to inclusion of
processes which are not present in the experiments.

To describe four-phonon decays in experimentally relevant nanotubes,
we carry out numerically the discrete triple sum over intermediate
momenta in Eq.~(\ref{Qinv4}) using the discrete frequencies of a
finite-length doubly clamped beam and experimentally observed inverse
lifetimes $\eta$. The result then gives the \emph{contribution} to the
observed losses from four-phonon processes.  The resulting
\emph{contributions} to the quality factor are $Q_4=6.6\times 10^8$ at
$T=1$~K and $Q_4=1.6\times 10^{14}$ at $T=0.01$~K for a tube of length
$L=800$~nm and $R=1.5$~nm \cite{vanderZant}.  This indicates that, at
low temperatures, four-phonon decays give only a small contribution to
the observed losses, which are $Q\approx 10^4$ at $T=1$~K and $Q
\approx 10^5$ at $T=0.01$~K \cite{vanderZant}.  Although small, these
losses are still much greater than the three-phonon contributions:
$Q_3=3.1\times 10^{27}$ at $1$~K and even greater at $0.01$~K. Our
analysis, contrary to the aforementioned theoretical studies, suggests
that the losses observed by H\"uttel et al.\cite{vanderZant} are
mostly extrinsic.

J.L. supported by DOE \# DE-FG02-07ER46432.

\bibliography{CNTpaper}
\end{document}